# A Bioinformatic Study of Genetics Involved in Determining Mild Traumatic Brain Injury Severity and Recovery

Mahnaz Tajik [a,b,c], Michael D Noseworthy [a,b,c,d,e,*]

a. School of Biomedical Engineering, McMaster University, 1280 Main St W, Hamilton, ON, Canada;
b. Imaging Research Centre, St. Joseph's Healthcare Hamilton, 50 Charlton Ave E, Hamilton, ON, Canada;
c. Department of Medical Sciences, McMaster University, 1280 Main St W, Hamilton, ON, Canada
d. Department of Electrical and Computer Engineering, McMaster University, 1280 Main St W, Hamilton, ON, Canada;
e. Department of Radiology, McMaster University, 1280 Main St W, Hamilton, ON, Canada

* Corresponding Author
Dr. Michael D Noseworthy, PhD, PEng
St Joseph's Healthcare Hamilton, Fontbonne Building Room 130A
50 Charlton Ave E, Hamilton, ON, Canada, L8N 4A6
+1 905.525.9140 x 23727
nosewor@mcmaster.ca

Mahnaz Tajik: tajikm1@mcmaster.ca
ORCID id: 0009-0006-2455-689X
Michael D Noseworthy: nosewor@mcmaster.ca
ORCID id: 0000-0003-1464-159X

**Authors Biography Note**

**Dr. Michael D. Noseworthy** is a professor of electrical and computer engineering, and he is special professional staff in Radiology and Nuclear Medicine at St. Joseph's Healthcare, Hamilton. He received M.Sc. and PhD degrees from the University of Guelph, specializing in applications of magnetic resonance imaging (MRI), nuclear magnetic resonance (NMR), biochemical assays, electron microscopy and electron paramagnetic resonance (EPR) methods to assess free radical induced liver and brain damage.

**Mahnaz Tajik** completed her B.S. in Cellular and Molecular Biology in Iran. She then earned an M.Sc. in Biomedical Engineering from McMaster University, where she is now a Ph.D. candidate in Medical Sciences.

## Abbreviations:

**(TBI)** Traumatic Brain Injury, **(mTBI)** Mild Traumatic Brain Injury, **(microRNA/ miRNA)** Micro Ribonucleic Acid, **(RNA-seq)** RNA-Sequencing, **(BBB)** Blood-Brain Barrier, **(GO)** Gene Ontology, **(KEGG)** Kyoto Encyclopedia of Genes and Genomes, **(PPI)** Protein-Protein Interactions, **(GEO)** Gene Expression Omnibus, **(RNA)** Ribonucleic acid, **(mRNA)** Messenger RNA, **(OMIM)** Online Mendelian Inheritance in Man, **(DAVID)** Database for Annotation, Visualization, and Integrated Discovery, **(MF)** Molecular function, **(BP)** Biological process, **(CC)** Cellular components, **(GSEA)** Gene Set Enrichment Analysis, **(BWA)** Burrows-Wheeler Alignment, **(MMA)** Mixed Martial Arts, **(hg38)** Human genom38, **(FDR)** False Discovery Rate, **(Padj)** P.adjusted, **(DAergic)** Dopaminergic, **(PTSD)** Post-traumatic stress disorder, **(UPP)** Ubiquitin-proteasome pathway, **(FHM)** Familial Hemiplegic Migraine, **(FTD)** Frontotemporal dementia, **(FADS1)** Fatty Acid Desaturase 1, **(SNP)** Single Nucleotide Polymorphism, **(CSF)** Cerebrospinal Fluid

Data availability: The coding and analysis details are available upon request. For further information, please contact the corresponding authors.

Funding statement: No specific funding was associated with this project.

## Abstract

Aim: This in silico study sought to identify specific biomarkers for mild traumatic brain injury (mTBI) through the analysis of publicly available gene and miRNA databases, hypothesizing their influence on neuronal structure, axonal integrity, and regeneration.

Methods: This study implemented a three-step process: (1) Data searching for mTBI-related genes in Gene and MalaCard databases and literature review ; (2) Data analysis involved performing functional annotation through GO and KEGG, identifying hub genes using Cytoscape, mapping protein-protein interactions via DAVID and STRING, and predicting miRNA targets using miRSystem, miRWalk2.0, and mirDIP (3) RNA-sequencing analysis applied to the mTBI dataset GSE123336.Results: Eleven candidate hub genes associated with mTBI outcome were identified: *APOE, S100B, GFAP, BDNF, AQP4*, *COMT*, *MBP*, *UCHL1*, *DRD2*, *ASIC1*, and *CACNA1A*. Enrichment analysis linked these genes to neuron projection regeneration and synaptic plasticity. miRNAs linked to the mTBI candidate genes were hsa-miR-9-5p, hsa-miR-204-5p, hsa-miR-1908-5p, hsa-miR-16-5p, hsa-miR-10a-5p, has-miR-218-5p, has-miR-34a-5p, and has-miR-199b-5p. The RNA sequencing revealed 2664 differentially expressed miRNAs post-mTBI, with 17 showing significant changes at the time of injury and 48 hours post-injury. Two miRNAs were positively correlated with direct head hits.

Conclusion: Our study indicates that specific genes and miRNAs, particularly hsa-miR-10a-5p, may influence mTBI outcomes. Our research may guide future mTBI diagnostics, emphasizing the need to measure and track these specific genes and miRNAs in diverse cohorts.

## Introduction

A traumatic brain injury (TBI) is a neurological injury that can lead to debilitating cognitive, emotional, and physical symptoms in the acute and chronic stages of recovery (1,2). The majority (~90%) of TBI are classified as mild (i.e., mTBI) (3), however the subjectivity surrounding grading injury severity and underreporting of head injuries remain substantial challenges (4–6). Most adults recover within 10-14 days (1,7,8); however, approximately 20% of adults who suffer an mTBI have symptoms lasting more than one month (1,9,10). It has been shown that older individuals with a history of brain injury may be at higher risk of developing neurodegenerative diseases including Alzheimer's disease (11–13).

The stress and strain from mechanical forces applied during an mTBI event can cause shearing and compression of brain structures (14). Those biomechanical forces initiate the primary injury phase, characterized by loss of axonal and blood-brain barrier (BBB) integrity, which then progresses into a secondary phase as ion imbalances and neuroinflammation gradually become exacerbated (15–18). The secondary injury phase involves a neurometabolic cascade that affects brain structure and function on a molecular level that can develop into excitotoxicity, oxidative stress, and cell death (18–21). Due to the unique nature of each mTBI highly sensitive and objective diagnostic tools are eminently required to aid in personalized assessments and treatments for the diverse clinical presentations associated with these injuries.

Numerous genes and molecular factors, such as *APOE* (neuronal repair), *BDNF* (synaptic plasticity), *IL-6, IL-10, TNFα* (neuroinflammation), *ENO2*, *UCHL1* (neuronal injury), *GFAP*, and *S100B* (glial activation, blood-brain barrier integrity), are implicated in mTBI (22–27). These genes play important roles in critical processes such as inflammation, neuronal injury, repair, and cognitive dysfunction, influencing both short- and long-term outcomes. Short-term brain health following a head trauma could be related to genetic variants that affect the severity of axonal damage, BBB disruption, inflammation, neuronal survival, and cognitive dysfunction (28,29). Furthermore, long-term outcome may be determined by genes involved in neuroplasticity and neuronal regeneration (29). Research highlights the significance of immune-related gene expression changes in concussion. Simpson et al. (2024) examined peripheral blood transcriptomes in concussed athletes, revealing an initial activation of the immune response post-injury, which was later followed by cytokine downregulation (30). As a result, identifying and characterizing the role of specific genes and micro ribonucleic acids (miRNAs) associated with mTBI could lead to the development of targeted interventions based on injury specific diagnostic biomarkers.

miRNAs are small non-coding regulatory RNAs that directly regulate gene expression by preventing or increasing the translation of target messenger RNA (mRNA) (31–33). The substantial influence of miRNAs on brain development and function makes them potentially useful biomarkers with high specificity for mTBI severity and prognosis (34–36). In a healthy person the BBB protects the brain by providing highly specific transportation of small molecules and macromolecules such as proteins (37). However, direct head injury can cause damage that allows brain specific miRNAs to cross the BBB through microvesicles, exosomes, and lipoprotein carriers, leading to their abnormal presence in peripheral circulation (36,38). Recent studies emphasize the role of miRNAs, particularly exosomal miRNAs, as non-invasive biomarkers for diagnosing and understanding TBI (39). Yang et al. (2024) identified miR-206 and miR-549a-3p as key markers for tracking neuronal damage and recovery, supporting the importance of circulating miRNAs in brain injury response (39). Feng et al. (2024) found that hsa-miR-122-5p and hsa-miR-193b-3p were notably elevated in the blood of TBI patients after treatment, showing a correlation with injury severity and microglial activation (40). These results underscore the value of miRNA profiling in tracking brain injury progression and understanding its mechanisms.

In the current study, we used a bioinformatics workflow to systematically identify genes, and their regulating miRNAs specifically associated with mTBI. Bioinformatics analysis tools have been widely used to detect genes, miRNAs, and functional pathways involved in the pathogenesis and progression of TBI (41–44). *In silico* studies (i.e., which rely on databases and pathway analysis software) provide valuable information about direct and indirect gene interactions to identify alterations present in specific genes and miRNAs. However, mTBI is a heterogeneous condition that affects numerous neurological processes making it challenging to identify accurate and specific biomarkers. Currently, there are no meaningful and specific biomarkers that can be utilized to predict clinical outcome of mTBI (42).

Taking all of the above into account, the purpose of our study was to identify potential hub genes and biological pathways associated with mTBI neurological sequelae using multiple online databases and pathway analyses. Online miRNA bioinformatics tools were then used to predict target miRNAs correlated with hub genes and the pathophysiological processes associated with mTBI. Additionally, we searched the Gene Expression Omnibus or GEO database for RNA-sequencing studies on acute mTBI (45) to analyze freely accessible data and compare it with our miRNA predictions. Our goal was to determine how acute mTBI affects temporal expression of circulating miRNAs and to survey the biological and neuronal pathways related to these miRNAs. It was hypothesized that several genes related to neuronal regeneration and healthy cognition would be identified as hub genes related to mTBIs, and that several miRNAs would be identified as potentially useful biomarkers.

## **Methods**

Our previously published literature review article (46) focused on estimating how many genes are associated with neurological structural and functional changes related to mTBIs, and which genes are more significantly altered post-injury **(Figure 1)**. These genes were chosen by functional similarity, biological pathways, gene ontology, and phenotype related to mTBI outcome. We reviewed the physiological, molecular, and omics changes that are present following a brain injury, particularly mTBI, as well as earlier research on gene and miRNA changes, polymorphism, and the effect of single nucleotide polymorphisms (SNPs) on mTBI patient recovery (46). We also reviewed epigenetic mechanisms that follow head trauma and their impacts on gene expression and neurological dysfunction. We included studies on human and mammalian animal models that examined changes in blood biomarker levels following mTBI and their correlation with neurological symptoms. The search was performed using primary databases such as PubMed. Search terms included combinations of 'mild traumatic brain injury' OR 'mTBI' AND 'genes' AND 'miRNA'. We focused on experimental studies that explored the genetic and molecular factors contributing to the pathophysiology and recovery of mTBI (43). Inclusion criteria for gene selection were based on the following: (1) functional similarity to established mTBI-related pathways (such as neuroinflammation, neuronal repair, and synaptic plasticity); (2) involvement in pertinent biological processes (identified through gene ontology terms associated with inflammation, glial activity, cognition, and axonal repair); and (3) their purported influence on mTBI outcomes (such as functional recovery, cognitive impairment, and long-term neurological effects). Gene ontology, biological pathways, functional similarities, and symptoms associated with mTBI outcomes were taken into consideration while choosing our gene set. Based on our initial search, inclusion criteria, prevalence of genes in literature, their involvement in key biological pathways, and their relevance to mTBI-related symptoms, 30 genes were identified as potentially playing a role in mTBI neuronal health and function and selected for further analysis in our study.

***Bioinformatic Databases***

A comprehensive and thorough search of online bioinformatics databases was conducted using the gene databases from the National Center for Biotechnology Information (NCBI) (47) and the human disease database MalaCards (48) to determine how many genes have been correlated with mTBI (**Figure 1**). This search involved the utilization of specific search terms (such as `gene database`, `disease database` and `miRNA prediction database`), filters, and criteria to ensure the retrieval of relevant and up-to-date information. Gene databases supply gene-specific connections in the nexus of map, sequence, expression, structure, function, citation, and homology data, and the identity of a gene can be determined by specifying the sequence, map position, or phenotypic characteristics (47). These gene identification numbers are utilized across all NCBI datasets and kept up to date via annotation changes (47). This database incorporates data and linkages to the Online Mendelian Inheritance in Man (OMIM) database (https://www.omim.org/), which is a continuously updated catalog of human genes, genetic disorders, and traits. "Mild traumatic brain injury" was the key phrase used to search in the gene database and the results were automatically displayed in tabular format. The results were sorted by relevance to the condition and gene weight and humans were selected as the primary organism. As determined by the gene dataset, 82 genes and 5 miRNAs are known to be involved with mTBI in humans a list of affiliated genes with mTBI was determined based on a Gene Weight calculation, which included multiple lines of evidence such as gene expression, protein clusters, and OMIM entries (47).

MalaCards is a comprehensive database of diseases and their annotations that creates an electronic card for each of the 16,919 human disorders by combining and mining 44 data sources (48). MalaCards contains disease-specific prioritized annotations, and inter-disease connections based on GeneCards, GeneDecks, and their search capabilities (48). We performed a MalaCards search to determine how many genes are associated with TBI, and 17 genes were identified as key. The relevance score for each was calculated by considering the significance of the many resources linking the gene to the condition (48) (https://www.malacards.org/).

***Data Analysis: Functional Enrichment and Pathway Analysis***

The Database for Annotation, Visualization, and Integrated Discovery (DAVID, version 6.8) was used to address functional annotation, visualization, and biological meanings behind the mTBI-associated genes with specific Gene Ontology (GO) (49,50) (https://david.ncifcrf.gov). The DAVID bioinformatics resources include a combined biological pool of knowledge and algorithms for systematically deriving the biological role of genes from an extensive list of genes and proteins (50). GO defines the interaction across genes by annotating and categorizing the molecular function (MF), biological process (BP), and cellular components (CC) linked with a gene product (51). This process allowed enrichment analysis of a gene collection (51), which was carried out in this study to reveal which MF, BP, and CC were disproportionately represented in our list of human genes related to mTBIs. The Kyoto Encyclopedia of Genes and Genomes (KEGG) database was then used to explore molecular interactions and to create a network of relationships between our candidate genes (52) (https://www.genome.jp/kegg/pathway.html). To better understand the biological significance of mTBI candidate genes, our gene list was uploaded into DAVID for enrichment analysis related to GO and KEGG terms. A filter of 0.05 was used as the significance cut-off value for fold enrichment analysis in the DAVID database. A Gene Set Enrichment Analysis (GSEA) (https://www.gsea-msigdb.org/gsea/index.jsp) was performed that focused on gene groups that shared a similar biological function, regulation, and chromosomal location (53). GSEA analysis was performed on our candidate genes to identify biological processes or pathways that may cause neurological perturbation following mTBI (**Figure 2**).

***Protein-Protein Interaction (PPI) Network and Hub Genes***

The Search Tool for the Retrieval of Interacting Genes (STRING, version 11.0) is an online database that offers access to experimentally determined and predicted protein-protein interaction (PPI) data (54) (https://string-db.org). PPI networks were built using STRING to investigate the functional interactions between the mTBI candidate genes and proteins that we identified using DAVID and GSEA. The parameter was set to have a medium confidence score of > 0.4. To visualize the network, Cytoscape (version 3.8.2) (55) with the CytoHubba plug-in (56) was used to perform a network analysis on the top ten candidate genes to indicate interactions based on 11 scoring or topological approaches (56) (**Figure 2**). We used the degree of interaction as the main measure of relevance in this study and identified the top ten mTBI genes that were used moving forward in this study.

***miRNA-target Gene Regulatory Network***

The role of miRNAs is to control gene expression by interacting with their target genes during the post-transcriptional phase. In our study, online tools were used to first identify miRNAs that regulate mTBI candidate genes and then secondly, we constructed miRNA-target gene regulatory networks using Cytoscape software (**Figure 2**). To predict which miRNAs, regulate mTBI candidate genes, various prediction tools were used including miRSystem, miRWalk2.0, and mirDIP. The miRSystem (57) (http://mirsystem.cgm.ntu.edu.tw/) integrates with seven well-known miRNA target gene prediction programs, and miRNAs were selected based on whether the total hit value was greater than or equal to one. The miRWalk2.0 (http://mirwalk.umm.uni-heidelberg.de/) (58) interacts with 12 different online databases to predict miRNA. With this approach miRNAs were chosen if they appeared in at least three out of the 12 miRWalk databases (miRMap (https://mirmap.ezlab.org/) (59), Targetscan (http://www.targetscan.org/) (60), and miRDB (http://www.mirdb.org) (61)). The miRDIP, another comprehensive database for predicting miRNAs, was used to identify miRNAs with very high scores that were directly linked to identified mTBI-related genes (http://ophid.utoronto.ca/mirDIP/) (62). The DIANA mirPath (v.3) (http://diana.imis.athena-innovation.gr/DianaTools/index.php) (63) and the miRNA pathway dictionary (miRPathDB version 2.0) (64) (https://mpd.bioinf.unisb.de/overview.html) were also utilized to more precisely identify miRNAs associated with the mTBI candidate genes based on biological pathways and gene ontology enrichment. The only variations between these prediction tools are the algorithms and statistical analyses that each database employs to categorize miRNAs, but by employing several prediction tools the consistent miRNA across multiple sources allowed for corroboration of results.

***RNA-sequencing Analysis: Analysis of a mTBI Dataset***

Given the recent evidence of miRNA responses in brain injury (65), we performed miRNA sequencing analysis to explore differential expression patterns and their association with injury severity and recovery. A small RNA sequencing study of human serum and saliva prior to, during and after amateur mixed martial arts (MMA) competitions (dataset name: GSE123336) (45) was downloaded from the GEO repository to apply the result of our miRNA predictions to a human mTBI dataset (http://www.ncbi.nlm.nih.gov/geo) (66) as of March 2023 (**Figure 3**). Institutional ethics approval was not required at our site for this study as data was downloaded from a publicly available dataset and participants from the original dataset provided informed consent at the time of their data collection (42). GEO is a global public database that stores and openly disseminates high-throughput functional genomics data from microarray, next-generation sequencing, and other sources that are contributed by the science community (66). The dataset was based on Illumina NextSeq 500 (*Homo sapiens*) and consists of a total of 218 samples (131 serum and 87 saliva samples) (45). In that study, raw data was obtained from serum samples that were collected at different time points in relation to the acute mTBI (1-week pre-injury = 7, 0 days pre-injury = 52 samples, 0 days post-injury = 52 samples, 2-3 days post-injury = 17 samples, 1-week post-injury = 3) (45). The original paper did not provide a clear explanation for the decrease in sample numbers after the immediate post-injury time points. However, this reduction could likely be due to common issues in longitudinal studies, such as participant dropout, logistical challenges, or the unavailability of participants for follow-up sample collection. The *fastq* files belonging to 131 serum samples were directly downloaded from the European Nucleotide Archive (ENA) database.

### *Data Processing*

The original *fastq* files related to the GSE123336 data were downloaded from European Nucleotide Archive (ENA) (https://www.ebi.ac.uk/ena/browser/home). As a first step, *fastq* data were checked using the FastQC tool (Babraham Bioinformatics) (**Figure 4**). The *fastq* files contained 3' Illumina adapters that were trimmed using Trimmomatic software (67) (http://www.usadellab.org/cms/?page=trimmomatic). The data utilized for the analysis was single end read (i.e., the sequencer reads DNA fragments from one end to the other). Command line-based scripts were used to specify the trimming stages to remove the adapters from sequences. Subsequently, two rounds of alignment were carried out using the Burrows-Wheeler Alignment (BWA) software, one against the reference genome »hg38« and the other against the mature miRNA sequences (https://bio-bwa.sourceforge.net) (68). Aligning to the reference genome is a crucial supplementary measure, as it can further enhance the count matrix generated from miRBase alignment (69). In cases where reads did not align with miRBase, they were aligned with the genomic coordinates of mature miRNA present in the reference genome (69).

The BWA index function was used to build a full index for the genome reference »hg38«. A mature Fasta miRNA sequence, downloaded from miRBase database version 22 (70,71), then the BWA aln algorithm were used to align short read sequence with the indexes. Samtools view and Samtools sort were used to create bam files and sorted bam files respectively (72). FeatureCounts were conducted for counting reads and building a count matrix and comparing aligned reads with miRNA transcripts downloaded from the miRBase database (has.gff 3) (https://www.mirbase.org/) (70,71). As a final step, R software and the Differential gene expression analysis, based on the negative binomial distribution (DESeq2) package, were employed to identify differential miRNA expressions following an acute mTBI at the various time points and relative to the numbers of hits to the head the athlete sustained. Comparisons were made between each post-injury time point (0 days post-injury, 2-3 days post-injury, and 1-week post-injury) and the pre-injury baseline (0 days pre-injury and 1 week pre-injury). The pre-injury samples of each participant were compared to their equivalent post-injury samples, rather than being combined across time points. This ensured that each subject served as their own control, reducing variability due to inter-individual differences. Statistical analysis was used to identify miRNAs with significantly altered levels, based on a P-adjusted value threshold of $\leq 0.05$ (73). A Bonferroni correction was used to adjust for the total number of miRNAs analyzed across different time points.

## Results

### *Identifying Candidate Genes Using Databases and Literature*

According to published research, neurotrophic, inflammatory and catecholamine genes are more often altered after brain trauma and serve as reliable predictors of injury severity and recovery (28). These genes have substantial influence on biological processes associated with acute mTBI outcomes. To identify genes involved with mTBI, peer-reviewed journal articles (46) and search of two databases (Gene and MalaCards) were conducted. From these searches, 129 genes were significantly correlated with mTBI (82 genes from Gene database, 17 genes from MalaCards, and 30 genes were identified from experimental studies). For a gene to meet study inclusion criteria function, expression in the blood circulation system, and most direct relationship to mTBI neuronal dysfunction and recovery were considered. Genes were excluded that are expressed only in brain tissue, inflammatory genes, and cytokines. We excluded genes exclusively expressed in brain tissue because the goal of our research was to identify circulating genes, particularly those involved in blood circulation, which could potentially act as *in vivo* and non-invasive biomarkers of mTBI-related dysfunction. However, we acknowledge that excluding brain-specific genes could limit the scope of our findings, particularly in relation to their direct involvement in brain pathology and recovery after mTBI. Inflammatory genes and cytokines, although crucial in the context of mTBI, were excluded due to their typical association with general inflammatory responses and immune system regulation. Since inflammation is a complex and multifaceted process, we chose to maintain a more specific focus on genes directly linked to neuronal dysfunction and recovery. Based on these criteria (i.e., neurological symptoms, cognitive impairment, recovery timeline and hub gene analysis), we were able to focus the mTBI list from 129 to 11 genes: *APOE* (Apolipoprotein E), *S100B* (s100 calcium binding protein B-), *GFAP* (Glial fibrillary acidic protein), *BDNF* (Brain-derived neurotrophic factor), *AQP4* (Aquaporin-4), *COMT* (Catechol-O-methyltransferase), *MBP* (Myelin basic protein), *UCHL1* (Ubiquitin C-terminal hydrolase L1), *DRD2* (Dopamine receptor D2), *ASIC1* (Acid-sensing ion channel 1), and *CACNA1A* (Calcium voltage-gated channel subunit alpha 1 A). These genes were determined based on their functional similarities, involvement in biological pathways, phenotypes, protein interactions, and expression related to mTBI outcome.

***Data Analysis: Functional and Pathway Enrichment Analyses***

Functional enrichment analysis was performed to identify GO terms and KEGG pathways for the 11 mTBI candidate genes (**Figure 4**). The top 5 enriched GO terms and 2 significant KEGG pathways for our candidate genes based on fold enrichment analysis shown in table 1 **(Table 1)**. The biological processes connected with mTBI genes were considerably rich in neuronal processes, and more specifically, neuron projection regeneration, which is related to the regrowth of axons or dendrites in response to their loss or damage (Fold enrichment = 381.6, P = 4.80E-03). The neuronal cell body has the highest concentration of cellular components associated with mTBI genes, and tau protein binding was found to be the most enriched molecular function associated with mTBI genes (Fold enrichment = 279, P = 6.50E-03). Based on fold enrichment analysis, the only two significantly enriched KEGG pathways were cocaine addiction (Fold-enrichment = 35.1; P = 4.90E-02) and dopaminergic synapse pathways (Fold-enrichment = 20.2; P = 6.80E-03). Furthermore, GSEA analysis showed the majority of mTBI candidate genes were involved in cognition, synaptic signaling, memory, and nervous system processes. According to the GSEA phenotype analysis, *APOE*, *BDNF*, *CACNA1A*, *COMT* and *UCHL1* were identified as the main genes associated with cognitive impairment (FDR q-value < 0.05) (**Table 2**).

### *Protein-Protein Interaction (PPI) Network and Hub Genes Analysis*

By submitting the candidate genes into STRING, PPIs associated with mTBI candidate genes were obtained (**Figure 4**). The hub genes with CytoHubba, and their degree of interaction displays 10 nodes and 24 edges (**Figure 5)**. The interaction network revealed that most candidate genes exhibited strong interactions with each other, and enrichment analysis outcomes for hub genes showed different neurological tasks and neurological pathways related to the mTBI hub genes. For instance, *ASIC1, APOE, S100B, COMT* and *DRD2* play essential roles in memory function; *APOE, UCHL1, S100B, DRD2, BDNF* and *GFAP* play key roles in neuron projection development; *UCHL1, S100B, COMT, DRD2* and *APOE* are essential genes related to behavior, and *APOE, S100B, DRD2* and *BDNF* play a significant role in the regulation of cell death (these processes were selected based on their higher node degree and FDR q-value) (**Figure 5**).

### *miRNA-Target Analysis*

Online databases were utilized to predict target miRNAs connected to the mTBI candidate genes. Over 100 miRNAs were associated with each gene according to miRSystem, miRwalk 2.0, and miRDip. To identify significant miRNA closely linked to mTBI genes the DIANA mirPath

v3.0 bioinformatic tool and miRpathDB v2.0 database were used. DIANA miRPath obtained miRNA genes targets and KEGG pathways related to gene function, while miRpathDB used GO to display molecular function, cellular component, and biological process associated with mTBI selected miRNAs and their related genes. The results showed that miRNAs of hsa-miR-9-5p, hsa-miR-204-5p, hsa-miR-1908-5p, hsa-miR-16-5p, hsa-miR-10a-5p, has-miR-218-5p, has-miR-34a-5p, and has-miR-199b-5p were highly related to the mTBI candidate genes and exhibited overlap in databases thus possibly indicating dysfunction following mTBI. Pathway analysis revealed that predicted miRNAs targets were mainly engaged in nervous system signaling, neuron projection and cell differentiation (**Figure 6; Table 3**). Because neuronal signal transmission is governed by transmembrane voltage differences (74) an ion imbalance following brain injury may affect neuronal communication and resultant neurological impairment. In the cortex, projection neuron extension or process axons to distant intracortical, subcortical, and sub-cerebral targets regulate sensory input, motor functions, and cognitive abilities (75) all of which could be disrupted following mTBI.

***Identification of Differential Expression miRNAs in GSE123336***

In the GSE123336 study, samples were also taken based on the frequency of blows to the head, in addition to times before and after MMA competition. In light of this, we established conditions for each status (time points and number of hits) and analyzed the data based on each condition. The DESeq2 package of RStudio was used to identify differentially expressed miRNAs after acute mTBI, and the results showed that the expression profile of 2664 miRNAs had changed. In the first condition, we used the P-adjusted ($P_{adj}$) value (i.e. Bonferroni corrected) to compare 0-d post-mTBI with normal (0-d pre & 1-week preinjury) to identify significant miRNAs ($P_{adj} \leq 0.05$).

We employed a specific significance threshold of Padj≤0.05 after applying the Bonferroni Correction, which effectively controlled the issue of multiple comparisons while maintaining transparency in our analysis. We found that expression levels of 17 miRNAs were considerably altered immediately after injury (i.e., day 0, post), with 14 miRNAs showing upregulation and three displaying downregulation (**Table 4; Figure 7**). First, hsa-miR-10a-5p was overexpressed in serum immediately after injury ($P_{adj}$ = 9.48E-06), After 2days post-mTBI, similar miRNAs also displayed differential expression (17 miRNAs were found to have changed, with 14 increasing and 3 showing decreasing). Based on the analysis, there were no discernible changes in miRNA levels one week after injury compared to controls.

The same analysis, based on numbers of head hits, demonstrated that participants who got over 20 hits to the head had altered levels of miRNA expression, which increased the expression of some miRNAs like hsa-miR-10b-5p ($P_{adj}$ =0.039), and hsa-miR-143-3p ($P_{adj}$ =0.0082) (**Figure 7**In terms of GO and pathway analysis, the majority of miRNAs identified were related to nervous system development, cell projection, neuronal projection, metabolic processes, and neuronal system functions (**Table 4**).

## Discussion

According to our data and pathway analyses, 11 genes were found to have a significant association with neurological function, and 8 miRNA showed strong interaction with those 11 candidate genes. The candidate genes were engaged in cell differentiation, nervous system development, synaptic development, generation of neurons, and cell death. The miRNA identified as closely linked to mTBI were hsa-miR-9-5p, hsa-miR-204-5p, hsa-miR-1908-5p, hsa-miR-16-5p, hsa-miR-10a-5p, has-miR-218-5p, has-miR-34a-5p, and has-miR-199b-5p. Based on the evidence obtained from the DAVID and GSEA analyses, the mTBI candidate genes were mainly enriched in neuronal projection regeneration, regulation of neuronal synaptic plasticity, cognitive function, memory, and behavior. The discovery of specific genes and miRNA associated with neurological damage may be used as biomarkers and aid in the diagnosis and treatment of mTBI. There are numerous complex pathophysiological changes that occur following an mTBI. Therefore, the identification of brain-specific miRNAs can help improve the understanding of molecular alterations associated with the prediction of an individual mTBI patient's outcome. This study used various bioinformatic analyses to determine hub genes and miRNAs related to brain damage and neuronal dysfunction.

To establish the differential expression of miRNAs after brain damage, we also analyzed the GSE123336 dataset from the GEO repository and compared it with the results of our miRNA predictions. The results indicated significant changes in miRNA expression levels after acute mTBI. The hsa-miR-10a-5p was one of the miRNAs that showed significant increase immediately after injury (i.e., 0-day post injury), and our analysis indicated that it was an important miRNA related to BDNF and regulation of the cell morphogenesis pathway. We anticipate that it is a critical miRNA associated with acute mTBI. Previous studies have indicated the relevance of miR-10a-5p in various brain disorders, such as depressive disorder (76) and Parkinson's disease (PD) (77). Nonetheless, there are limited investigations that study the effect of this miRNA on the results of mTBI. Moreover, there was a considerable rise in hsa-miR-10b-5p and hsa-miR-143-3p at 0 days after injury, and there was a positive correlation between these changes and number of hits to the head, which meant that these miRNAs could serve as a useful biomarker to measure the severity of the direct head injury. Although further investigation is necessary to confirm and validate this hypothesis.

Previous research has shown that the majority of mTBI patients will recover from neurological dysfunction. However, as many as 15-30% will experience long-term neurocognitive and behavioral changes (78). Our findings suggest that mTBI candidate genes such as *APOE, S100B, GFAP, BDNF, AQP4, COMT, MBP, UCHL1, DRD2, ASIC1,* and *CACNA1A* may have an essential role in lasting mTBI-related neurological disorders. As determined by the GO analysis, *GFAP* and *APOE* were the two main genes involved in neuron regrowth, *S100B* and *APOE* were involved in the regulation of neuronal synaptic plasticity, *CACNA1A, DRD2,* and *UCHL1* would affect adult walking behaviour, and *ASIC1* and *DRD2* would contribute to learning. Further research is required to determine the association of these genes with recovery of neurological dysfunction after mTBI.

Our findings suggest that mTBI-specific miRNAs and genes are involved in critical processes such as neuronal repair, and cognitive dysfunction, which are likely to influence both short- and long-term neurological outcomes. These results align with prior research which identified 21 miRNAs that exhibited notable expression changes after mTBI and were linked to cognitive deficits and balance disturbances (42). They observed that miRNAs were more predictive of mTBI than conventional protein biomarkers like *GFAP* and *UCHL1*, which is consistent with our bioinformatics findings. Although we identified these proteins as part of the mTBI response, our focus on miRNA-gene interactions suggests that miRNAs may be more sensitive and specific indicators of injury and recovery.

In addition to the 11 candidate genes, our study determined 8 miRNAs to be associated with the mTBI candidate genes and neurological function. Based on computational analyses, these miRNAs might be involved in affecting mTBI pathophysiology. Previous studies indicated that some of these miRNAs were associated with brain disorders and suggested that they could be used as surrogate biomarkers (79). Based on predictions made through a bioinformatic study of miRNA with APOE as a regulatory target, miR-1908-5p was found correlated with genes involved in bipolar disorder (79). The location of miR-1908-5p is in the first intron of the fatty acid desaturase-1 (FADS1) gene on chromosome 11 (79).

Results from pathway analysis suggested that miR-1908-5p contributes to nervous system development and neuron projection. In the GENFI cohort, miR-204-5p was shown to be significantly lower in symptomatic frontotemporal dementia (FTD) compared to pre-symptomatic mutation carriers (80). Guedes *et al.*, found that miR-204-5p was upregulated in individuals with post-traumatic stress disorder (PTSD) compared to healthy controls (81). Moreover, Weisz *et al.*, found a substantial downregulation of miR-204-5p in chronic TBI patients (82). Surprisingly, miR-204-5p was identified to be abundant in brain, based on tissue expression patterns, and it also showed a dramatic increase 1 hour following concussion (83). These findings suggest that this miRNA could be used as a biomarker for neuropathological disorders. MiR-9 is another interesting miRNA that has attracted attention (84). It has a unique expression pattern in the brain and implements activities involved in central nervous system development (84,85). Brain development studies indicated that MiR-9 was related to gene networks that control the proliferation of neural progenitors (85). Therefore, the presence of this miRNA in neurological diseases is not surprising. For instance, miR-9 was found to be downregulated in Alzheimer's disease patients (86). Another significant miRNA we identified was miR-16-5p, which was previously shown as a viable potential biomarker for TBI, with the ability to distinguish between mild and severe cases (43).

Based on GO analysis miR-16 was involved in a variety of regulatory processes that are triggered by brain injuries, including positive apoptotic regulation via *BCL-2* targeting (87). Sun *et al.*, observed downregulation of miR-16-5p plays an important role in the faster recovery process in TBI patients via stimulation of osteoblast proliferation and the prevention of apoptosis (88). Researchers revealed that within the first 24 hours after mTBI, the level of miR-16-5p was significantly higher in mTBI patients compared to severe TBI patients (89). Furthermore, a recent study demonstrated that miR-16-5p and miR-21-5p are significantly upregulated in critical-fatal TBI cases, with miR-21-5p being particularly elevated in short-survival groups (90).

Also, RNA-seq analysis showed that miR-10a-5p, miR-10b-5p, and miR-143-3p levels in serum elevated immediately after mTBI. These results indicate that miR-10a-5p and miR-10b-5p are prominent targets for *BDNF*. Our analysis also shows a direct correlation between miR-10b-5p and miR-143-3p with the number of hits to the head. Therefore, these miRNAs may be useful candidate biomarkers to determine the severity of brain injury.

Enrichment analysis showed that miR-10a-5p and miR-10b-5p plays a significant role in neuronal processes when it comes to regulating cell morphogenesis and neuron generation, while miR-143-3p plays an active role in cell projection and cell-cell signaling. To our knowledge, there has not been a significant amount of research on miR-10a-5p and brain injury to recommend it as a potential biomarker. Previous studies have established that miR-143-3p is linked to ischemic stroke (91), and that miR-10b-5p is considerably differently expressed in Huntington's disease (92). Additionally, bioinformatic analyses revealed that miR-34a-5p miR-218-5p, and miR-199b-5p were targeted for their relationships to mTBI candidate genes, and these miRNAs are involved in pathways such as cell-cell signaling, cell death, regulation of metabolic process, cell differentiation, nervous system development.) (39)However, more research is needed to confirm the link between these miRNAs and post-mTBI symptoms.

While our study provides valuable insights into gene and miRNA alterations following mTBI, recognition of its limitations is crucial for a more comprehensive interpretation. We used bioinformatics tools and online databases to predict biomarkers that could be altered in mTBI. However, it is important to note that the accuracy and completeness of the available data as well as the reliability of the databases used may have introduced potential inaccuracies or variations in our findings. Furthermore, the dynamic nature of bioinformatics databases and tools indicates that changes over time could affect the relevance and accuracy of our results. Brain-specific genes, especially those expressed in neuronal and glial cells, play crucial roles in the underlying mechanisms of traumatic brain injury.

Disruption of the BBB following mTBI could allow some of these brain-specific molecules to circulate in the bloodstream, making them relevant targets for biomarker discovery. As such, we propose that future research should investigate both circulating and brain-specific genes to assess their combined potential in predicting mTBI severity and recovery. Another limitation of our research was the inability to access patients and biological samples, such as blood, to validate the predicted biomarkers. This limitation restricts the generalizability of our findings to patients with truly acute or chronic mTBI. Collecting samples from mTBI patients at various time points would improve our validity. We were unable to include a control group for circulating biomarkers, particularly miRNAs. Introducing controls from patients with extracranial injuries (e.g., orthopedic patients) in future studies could enhance the specificity of the identified biomarkers, such as hsa-miR-10a-5p. This would help determine whether extracranial injuries impact the levels of candidate miRNAs, which could affect their reliability as diagnostic tools for mTBI. The reduced number of samples at 2-3 days post-injury (17 samples) and 1-week post-injury (3 samples) compared to 0 days post-injury (52 samples) was another limitation of our study. This decrease in sample size at later time points may have been due to logistical challenges and participant availability for follow-up assessments, which is common in longitudinal studies. We acknowledge this limitation and recommend that future studies with larger sample sizes across all time points would enable a more comprehensive analysis of time-dependent biomarker trajectories. We were unable to differentiate between acute and chronic mTBI cases. This limitation arose from our reliance on databases and *in silico* analysis during our research. Lastly, for future studies, integrating bulk RNA-seq with single-cell transcriptomics could offer a more precise view of neuroplasticity and neuronal repair in mTBI recovery. Specifically, exploring DLK inhibition and ATF3 modulation as therapeutic targets may help mitigate neuronal loss (93).

## **Conclusions**

Our study identified 11 genes and 8 miRNAs that may be closely associated with mTBI severity and recovery outcomes. They were mainly involved in neurite regeneration, nervous system signaling, neurite outgrowth, and cell differentiation. The result of our study provides direction for potential gene and miRNA probes that could be developed into biomarkers for acute mTBI. Recent advances in research have expanded the understanding of miRNA biomarkers in mTBI, demonstrating their potential for diagnosis, prognosis, and treatment development (35,40,65,94). By integrating these findings, our study suggests that circulating miRNAs, particularly those involved in neuronal repair, could serve as non-invasive biomarkers for classifying mTBI severity and predicting long-term recovery.

Our study predicts key genes and miRNAs linked to mTBI severity and recovery, highlighting their role in neuronal plasticity and neuronal repair, while Jia et al. demonstrate that microglial activation protects against white matter atrophy, whereas endothelial dysfunction worsens neurodegeneration via BBB disruption (95), underscoring the need for targeted therapies integrating miRNA biomarkers, neuroimaging, and inflammatory markers. We anticipate that once these biomarkers are validated and a standardized analytical protocol is established, patients with mTBI will benefit from more precise assessments, targeted treatment strategies, and improved recovery predictions.

**Data Availability Statement** The data that support the findings of this study are openly available in Gene Expression Omnibus (GEO) (https://www.ncbi.nlm.nih.gov/geo/), reference number is GSE123336.

## Acknowledgements

Funding: No specific funding was associated with this project.

Conflict of interest: Dr. Noseworthy is the co-founder and of TBIFinder Inc. a data analytics company focused on MRI data, and this has no relationship to the work in this current study.

Author contributions: **Mahnaz Tajik:** Conceptualization, Methodology, Software, Validation, Formal Analysis, Investigation, Data Curation, Writing – Original Draft, Writing – Review & Editing, Visualization **Michael D Noseworthy:** Conceptualization, Resources, Writing – Review & Editing, Visualization, Supervision, Project Administration.

## Figures Captions

**Figure 1.** Step one of the methodologies was to explore mTBI related genes in the databases (Gene database from National Center for Biotechnology Information (NCBI) and MalaCard) and a comprehensive literature review of genetic focused mTBI research.

**Figure 2**. Step two of the methodology was performed after investigating genes using Gene Database and Malacards, the DAVID and GSEA databases were used to find relevant genes and biological pathways. Based on the enrichment analysis, 11 genes out of 129 exhibited a significant relationship with mTBI outcomes in both datasets. Protein-protein interaction and network analyse were performed utilizing string databases, while 10 Hub genes were identified using Cytohubba. Multiple miRNA databases were used to predict miRNAs linked to mTBI candidate genes and key pathways tied to mTBI changes.

**Figure 3.** Step three of the methodology was to investigate the GEO repository and download RNA-seq data associated with mTBI for mapping and differential expression analysis.

**Figure 4**. (A) Potential genes related to mTBIs in humans based on our literature review and database searches. These genes were filtered based on neural functions, and purple nodes show the genes that have more neurological functions. The protein-protein interaction network was visualized using Cytoscape v.3.8.2. (92 nodes and 327 edges). (B) mTBI candidate genes according to the enrichment analysis and PPI and co-expression between them indicated by Cytoscape v.3.8.2. The boldness of the arrows represents the strength of the interaction between candidate genes.

**Figure 5**. The PPI network of the 10 Hub genes is clustered by CytoHubba in cytoscape software based on the degree of interactions. The degree is indicated with the color of nodes, where darker colors represent a higher degree of interaction, and lighter colors indicate a lower degree of interaction. Based on the CytoHubba analysis BDNF showed the highest degree of interaction, while CACNA1A showed a lower degree of interaction.

**Figure 6**.  The miRNA targets' regulatory network. Predicted miRNA targets are displayed in green boxes, and purple lines indicate their connections to mTBI candidate genes (red ovals). Pathway analysis revealed that these miRNAs targets were mainly engaged in nervous system signaling, neuron projection and cell differentiation.

**Figure 7.**  The miRNAs were determined to be related to mTBIs and their predicted targets. These miRNAs show differential expression based on the BWA analysis.  The mirsystem, mirwalk databases were used to predict target genes. miRNAs are shown with a red diamond, and genes are shown with a purple circle.

## Tables:

**Table 1**. DAVID functional analysis for the 11 mTBI genes. The highest fold enrichment analysis represents GO terms that are strongly related to candidate genes. (Abbreviations: Biological Process (BP), Cellular component (CC), Molecular Function (MF)).

| **Category** | **GO term** | **Count** | **Fold Enrichment** | **P-Value** |
|---|---|---|---|---|
| GOTERM_BP_DIRECT | GO:0031102- neuron projection regeneration | 2 | 381.6 | 4.80E-03 |
| GOTERM_BP_DIRECT | GO:0048168- regulation of neuronal synaptic plasticity | 2 | 190.8 | 9.50E-03 |
| GOTERM_BP_DIRECT | GO:0007628- adult walking behavior | 3 | 147.7 | 1.50E-04 |
| GOTERM_BP_DIRECT | GO:0008306- associative learning | 2 | 127.2 | 1.40E-02 |
| GOTERM_BP_DIRECT | GO:0043407- negative regulation of MAP kinase activity | 2 | 84.8 | 2.10E-02 |
| GOTERM_CC_DIRECT | GO:0044297- cell body | 2 | 52.6 | 3.40E-02 |
| GOTERM_CC_DIRECT | GO:0043025- neuronal cell body | 5 | 26.3 | 1.70E-05 |
| GOTERM_CC_DIRECT | GO:0030425- dendrite | 3 | 14.8 | 1.40E-02 |
| GOTERM_CC_DIRECT | GO:0005886- plasma membrane | 8 | 3.2 | 6.10E-02 |
| GOTERM_CC_DIRECT | GO:0005737- cytoplasm | 7 | 2.2 | 3.80E-02 |
| GOTERM_MF_DIRECT | GO:0048156- tau protein binding | 2 | 279 | 6.50E-03 |
| GOTERM_MF_DIRECT | GO:0042802- identical protein binding | 4 | 8.2 | 8.30E-03 |
| KEGG_PATHWAY | hsa05030 - Cocaine addiction | 2 | 35.1 | 4.90E-02 |
| KEGG_PATHWAY | hsa04728 - Dopaminergic synapse | 3 | 20.2 | 6.80E-03 |

**Table 2**. Gene set enrichment analysis for mTBI candidate genes. A GSEA analysis revealed that most mTBI genes involved cognitive, synaptic signaling, memory, and nervous system functions. According to this analysis, APOE, BDNF, CACNA1A, COMT, and UCHL1 are the major genes associated with cognitive impairment (P-value = 3.40E-07). (Abbreviations: Biological Process (BP), Cellular component (CC), Human phenotype (HP)).

| **GO Terms** | **Genes in Overlap** | **P-value** | **FDR p-value** |
|---|---|---|---|
| GOBP_COGNITION | APOE, BDNF, DRD2, ASIC1, S100B | 6.48E-09 | 9.05E-05 |
| GOBP_SYNSPTIC_SIGNALING | APOE, BDNF, DRD2, ASIC1, MBP, CACNA1A | 1.27E-08 | 9.05E-05 |
| GOBP_MEMORY | APOE, BDNF, DRD2, ASIC1 | 1.53E-08 | 9.05E-05 |
| GOBP_NERVOUS_SYSTEM_PROCESS | APOE, BDNF, DRD2, ASIC1, S100B, MBP, AQP4 | 1.91E-08 | 9.05E-05 |
| GOBP_REGULATION_OF_TRANS_SYNAPTIC_SIGNALING | APOE, BDNF, DRD2, ASIC1, AQP4 | 4.44E-08 | 1.40E-04 |
| GOBP_BHAVIOR | APOE, BDNF, DRD2, ASIC1, S100B | 1.85E-07 | 4.99E-04 |
| GOCC_SOMATODENTRIC_COMPARTMENT | APOE, BDNF, DRD2, MBP, CACNA1A, COMT | 3.10E-08 | 1.17E-04 |
| GOCC_SYNAPSE | APOE, BDNF, DRD2, ASIC1, MBP, CACNA1A | 4.14E-07 | 8.70E-04 |
| HP_COGNITTIVE_IMPAIRMENT | APOE, BDNF, CACNA1A, COMT, UCHL1 | 3.40E-07 | 8.04E-04 |
| HP_MENTAL_DERERIORATION | APOE, CACNA1A, COMT, UCHL1 | 1.29E-06 | 2.45E-03 |

**Table 3.** Target miRNAs related to mTBI candidate genes. Canonical pathways selected by P-value < 0.05 as the significance cut-off value for miRNA pathway dictionary (miRPathDB version 2.0).

| Target miRNAs | Genes | Canonical pathways |
|---|---|---|
| hsa-miR-9-5p | DRD2, AQP4, MBP, BDNF, | Cell differentiation, Synaptic development, Regulation of cell differentiation |
| hsa-miR-204-5p | BDNF | Regulation of cell differentiation, generation of neurons |
| hsa-miR-1908-5p | APOE | Nervous system development, Synapse |
| hsa-miR-16-5p | BDNF, APOE, GFAP, COMT, MBP, ASIC1 | Synapse, Regulation of protein metabolism, Regulation of metabolism process |
| hsa-miR-10a-5p | BDNF | Regulation of cell morphogenesis, Generation of neurons |
| has-miR-218-5p | UCHL1 | Regulation of metabolic process, Cell differentiation, Nervous system development |
| has-miR-34a-5p | GFAP, BDNF, APOE, CACNA1A, ASIC1 | Cell-Cell signaling, Cell death |
| has-miR-199b-5p | COMT | Synapse, Synaptic membrane |

**Table 4.** The level of miRNAs that were differentially expressed at zero days and 2-3 days after injury. The miRwalk and mirsystem databases were used to investigate predicted targets, while mirpathDB was used to analyze pathways. All of these are significant based on $P_{adj} \leq 0.05$, some with very high significance.

| miRNAs ID (0 d post) | P value | $P_{adj}$ | Predicted Targets | Pathways |
|---|---|---|---|---|
| hsa-miR-145-3p | 2.06E-04 | 0.0176 | BDNF | Generic Transcription Pathway, Gene expression (Transcription), regulation of metabolic process |
| hsa-miR-873-3p | 1.12E-04 | 0.0118 | S100B | There is no notable pathway. |
| hsa-miR-125b-2-3p | 2.86E-08 | 9.78E-06 | GFAP, BDNF, AQP4, DRD2, COMT, UCHL1 | Synapse, regulation of nitrogen compound metabolic process, nervous system development |
| hsa-miR-99a-5p | 3.61E-06 | 5.92E-04 | There are no related genes | There is no notable pathway. |
| hsa-miR-143-3p | 8.95E-11 | 1.22E-07 | CACNA1A | Metal ion binding, cell projection, regulation of signaling, cell-cell signaling |
| hsa-miR-10b-5p | 2.22E-09 | 1.51E-06 | BDNF | Regulation of cell morphogenesis, regulation of cellular component organization |
| hsa-miR-10a-5p | 2.08E-08 | 9.48E-06 | ASIC1, BDNF | Regulation of cell morphogenesis, anatomical structure morphogenesis |
| hsa-miR-192-5p | 1.89E-06 | 3.69E-04 | BDNF, ASIC1, GFAP | Regulation of cellular process, regulation of metabolic process |
| hsa-miR-378a-3p | 1.33E-05 | 1.81E-03 | AQP4, BDNF, GFAP | Nervous system development, cell morphogenesis, regulation of cell morphogenesis, regulation of cellular component organization |
| hsa-miR-99b-5p | 2.79E-07 | 7.63E-05 | DRD2, COMT | Metabolic process, protein binding |
| hsa-miR-125a-5p | 7.39E-07 | 1.68E-04 | DRD2, CACNA1A, ASIC1, GFAP | Synapsen, negative regulation of signaling, intrinsic component of membrane |
| hsa-miR-24-3p | 3.78E-05 | 4.70E-03 | UCHL1, GFAP, MBP, S100B, DRD2, COMT, ASIC1 | Regulation of cellular component organization, regulation of autophagy, nervous system development, neurogenesis, synapse, |
| hsa-miR-345-5p | 3.47E-04 | 0.0278 | GFAP, BDNF, S100B, DRD2, ASIC1, COMT | Intracellular signal transduction, neuron differentiation, synapse, cellular localization, negative regulation of signaling, neuron projection, postsynapse |
| hsa-miR-27b-3p | 7.18E-05 | 8.17E-03 | BDNF, DRD2, CACNA1A, ASIC1 | Regulation of cell projection organization, Neuronal System, chemical synaptic transmission, metal ion binding, neurotransmitter secretion, synapse |
| hsa-miR-191-5p | 5.16E-04 | 0.0391 | GFAP, BDNF | Regulation of nitrogen compound metabolic process |
| hsa-miR-26a-5p | 0.000142 | 0.0139 | BDNF, S100B | Gene expression, Generic Transcription Pathway, regulation of cellular component organization, nervous system development |
| hsa-miR-184 | 3.90E-06 | 5.92E-04 | AQP4 | Cell projection, plasma membrane, plasma membrane bounded cell projection |

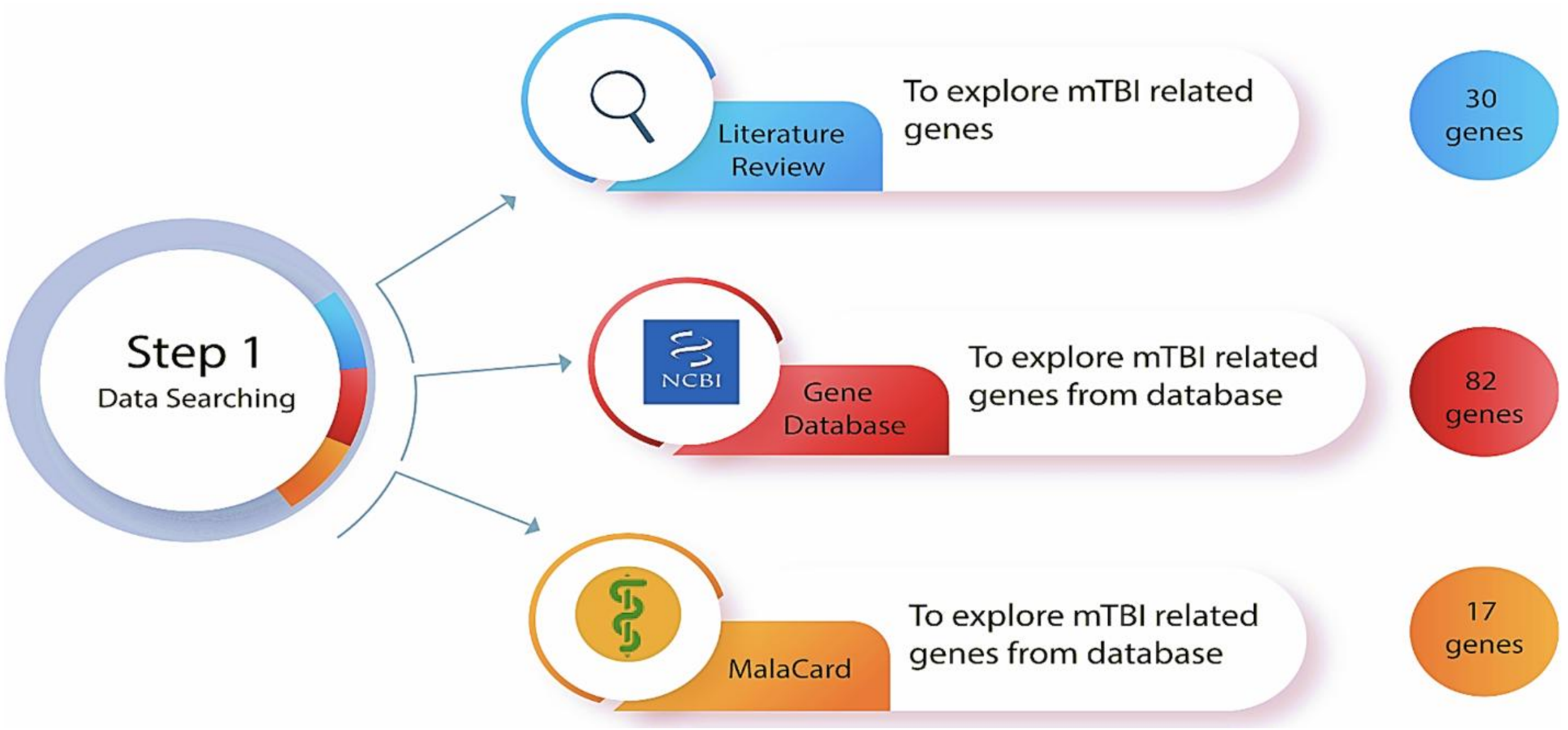

Figure 1.

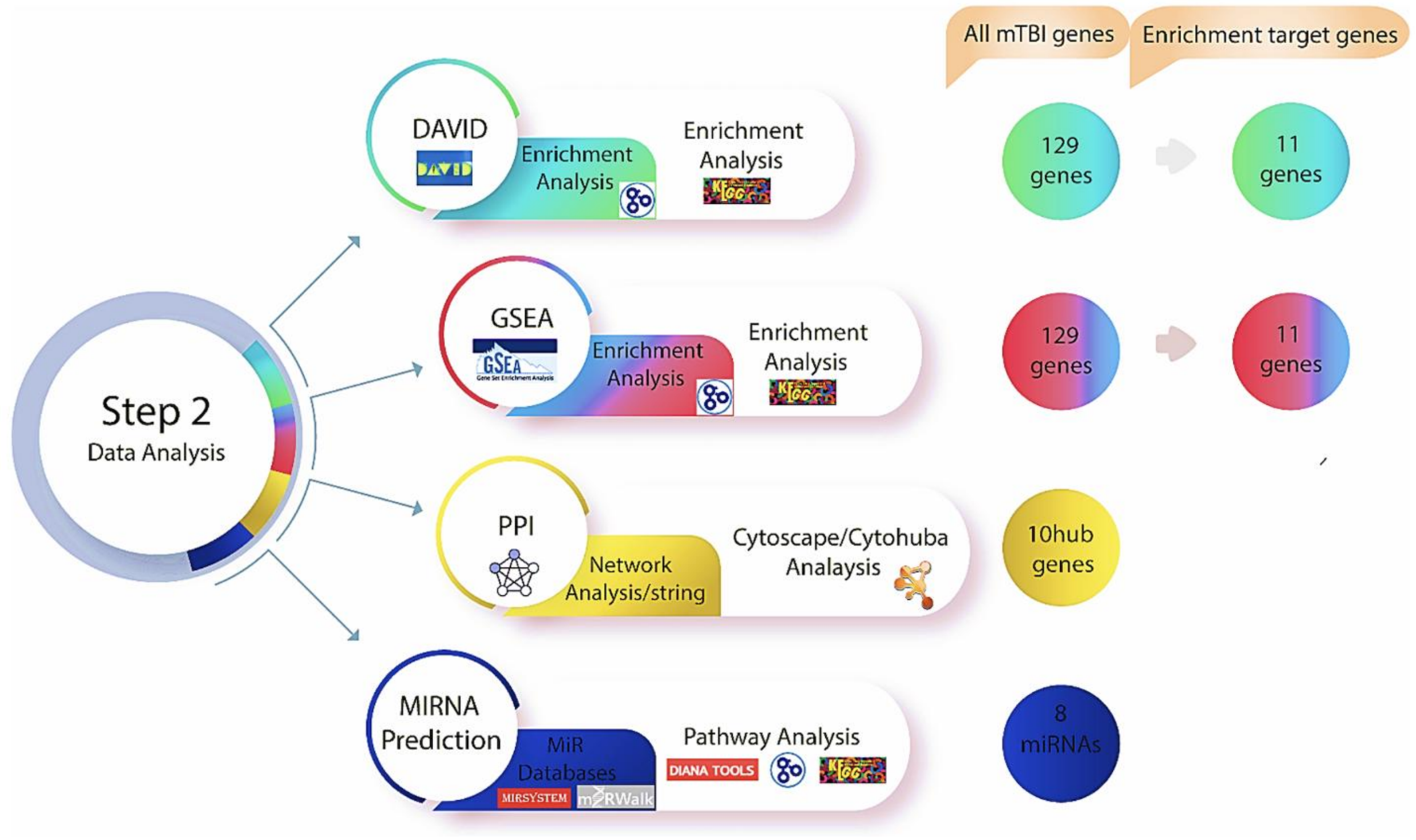

Figure 2.

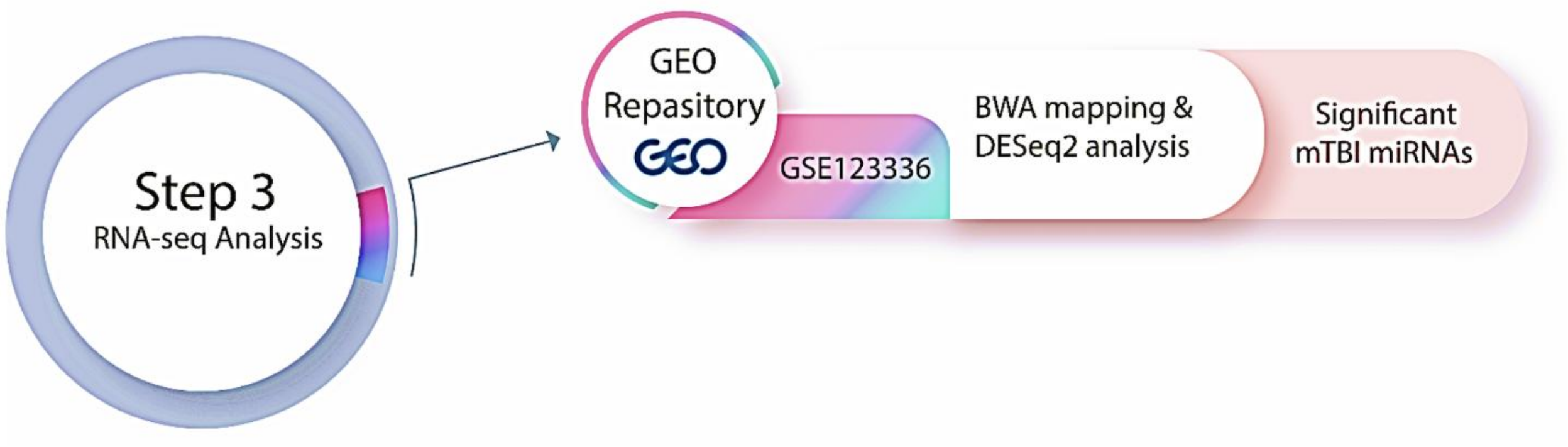

Figure 3.

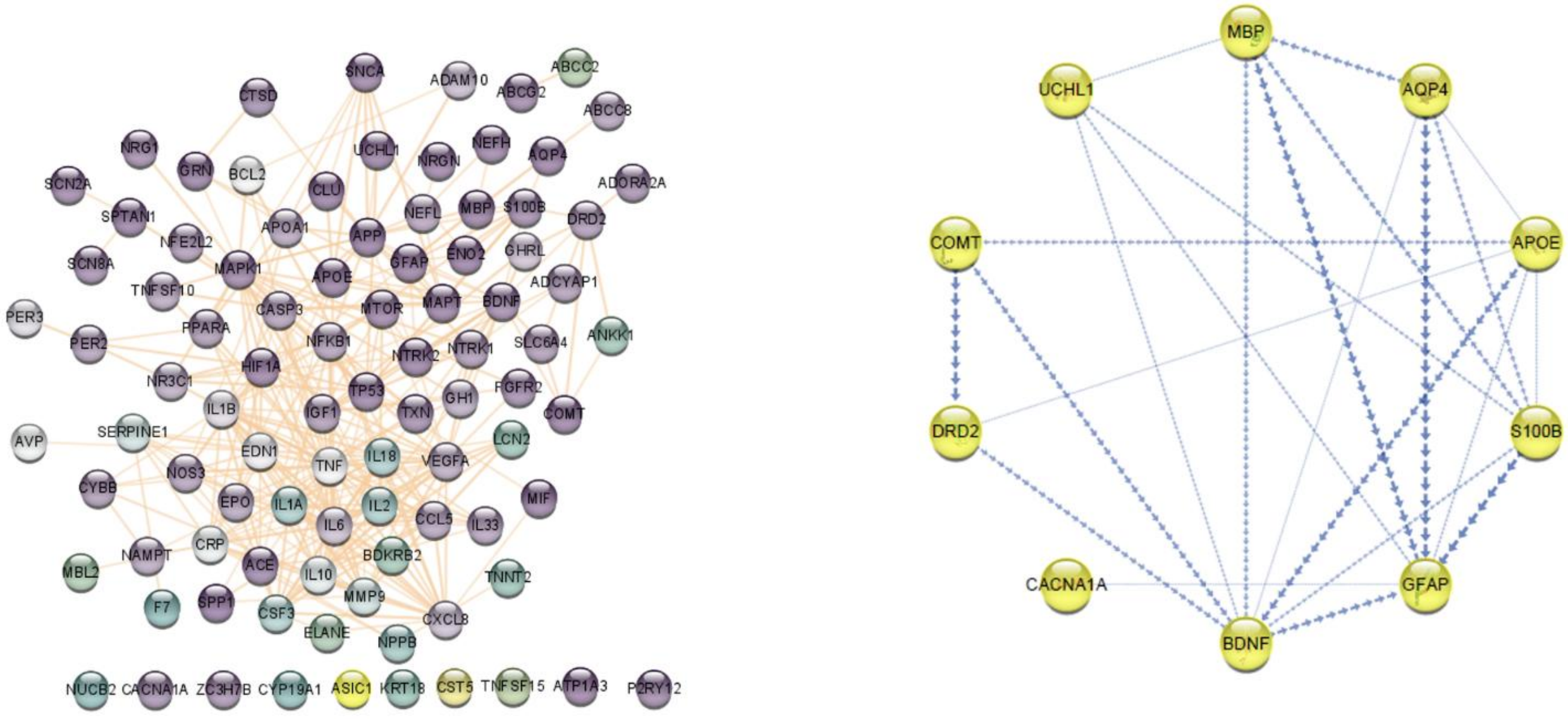

(A) (B)

Figure 4.

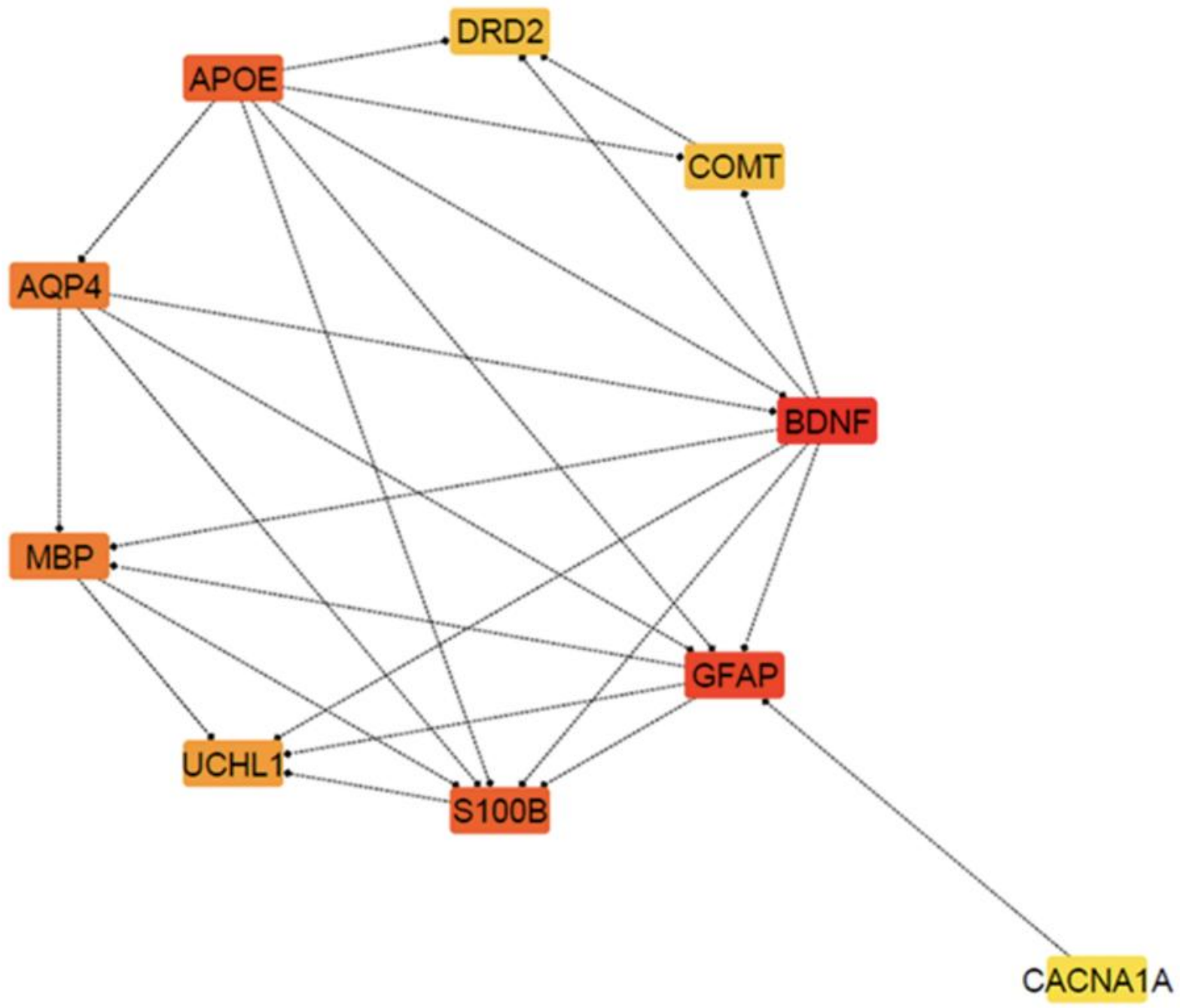

| Rank | Name | Score |
|---|---|---|
| 1 | BDNF | 8 |
| 2 | GFAP | 7 |
| 3 | APOE | 6 |
| 3 | S100B | 6 |
| 5 | MBP | 5 |
| 5 | AQP4 | 5 |
| 7 | UCHL1 | 4 |
| 8 | COMT | 3 |
| 8 | DRD2 | 3 |
| 10 | CACNA1A | 1 |

Figure 5.

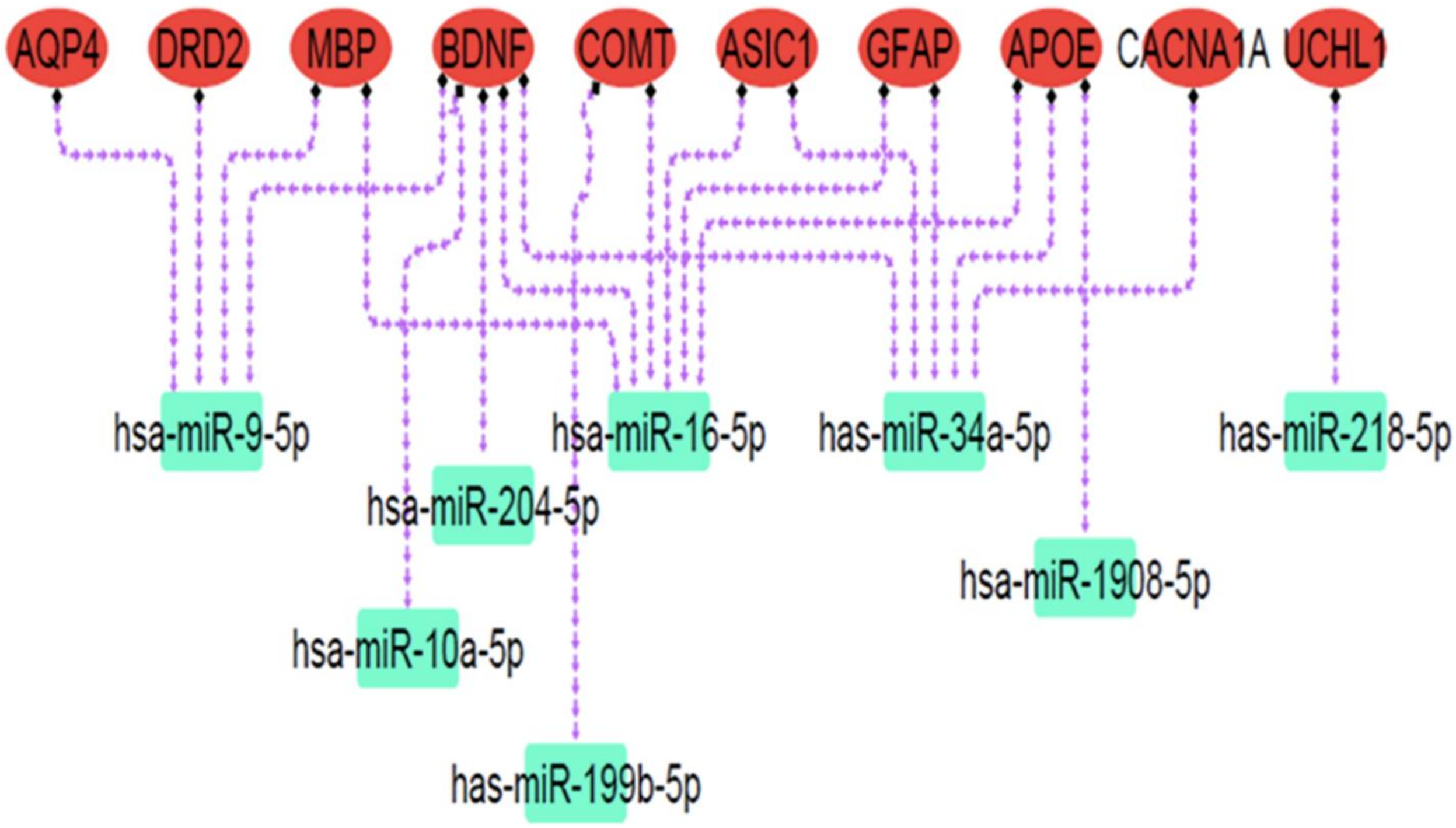

Figure 6.

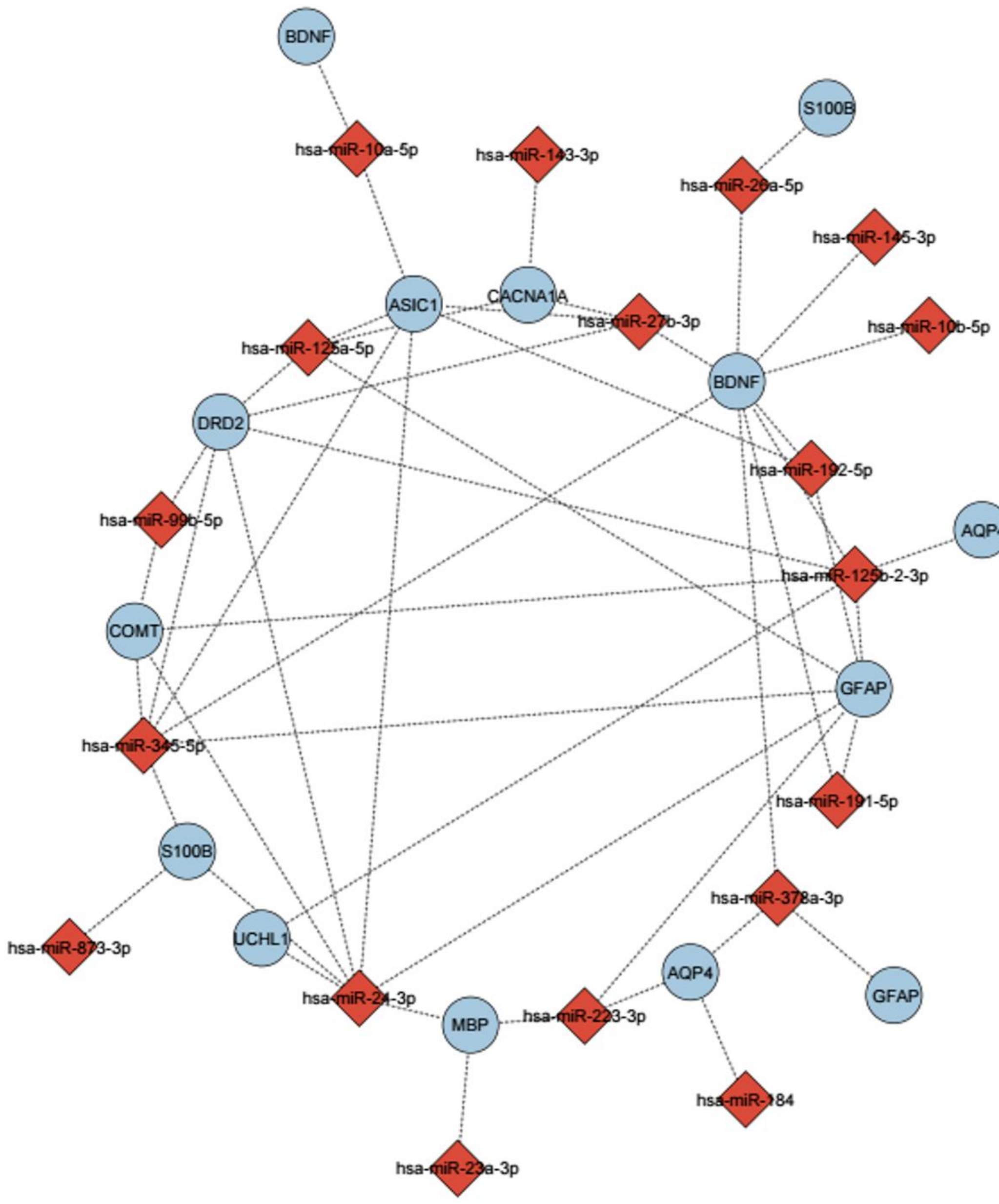

Figure 7.